\documentclass{emulateapj}
\usepackage{psfig,times}

\newcommand{\chandra}{{\em Chandra}}

\newcommand\xmm{{\em XMM-Newton}}

\newcommand{\kms}{{km\,s$^{-1}$}}
\newcommand{\net}{{$n_{\rm e}t$}}
\newcommand{\mug}{{$\mu$G}}
\newcommand\netunit{cm$^{-3}$s}

\newcommand\rcw{{RCW 86}}

\begin{document}

\title{The X-ray synchrotron emission of RCW 86 
and the implications for its age}

\shortauthors{J. Vink et al.}

\author{Jacco Vink\altaffilmark{1,2},Johan Bleeker\altaffilmark{1,2}, 
Kurt van der Heyden\altaffilmark{3}, Andrei Bykov\altaffilmark{4}, 
Aya Bamba\altaffilmark{5},
Ryo Yamazaki\altaffilmark{6}}

\email {j.vink@astro.uu.nl}
\altaffiltext{1}{Astronomical Institute, University Utrecht, P.O. Box 80000, 
3508TA Utrecht, 
The Netherlands}
\altaffiltext{2}{SRON Netherlands Institute for Space Research, Sorbonnelaan 2,
3584CA, Utrecht, The Netherlands}
\altaffiltext{3}{South African Astronomical Observatory
P.O. Box 9, Observatory 7935,South Africa}
\altaffiltext{4}{Ioffe Physico-Technical Institute,
Politekhnicheskaya 26, 194021 St.Petersburg,Russia}
\altaffiltext{5}{RIKEN, 2-1, Hirosawa, Wako, Saitama 351-0198, Japan}
\altaffiltext{6}{Department of Physics, Hiroshima University, 
Higashi-Hiroshima, 739-8526, Japan}

\begin{abstract}
We report here X-ray imaging spectroscopy observations of the northeastern
shell of the supernova remnant \rcw\ with \chandra\ and \xmm.
Along this part of the shell the dominant X-ray radiation mechanism
changes from thermal to synchrotron emission. We argue that both the
presence of X-ray synchrotron radiation and the width of the synchrotron
emitting region suggest a locally higher shock velocity of 
$V_s \approx 2700$~\kms\ and a magnetic field of $B \approx 24\pm 5~\mu$G.
Moreover, we also show that a simple power law cosmic ray electron spectrum
with an exponential cut-off cannot explain the broad band synchrotron emission.
Instead a concave electron spectrum is needed, 
as predicted by non-linear shock acceleration models.
Finally, we show that the derived shock velocity strengthens the case that
RCW 86 is the remnant of SN 185.
\end{abstract}


\begin{figure*}
\centerline{
 \psfig{figure=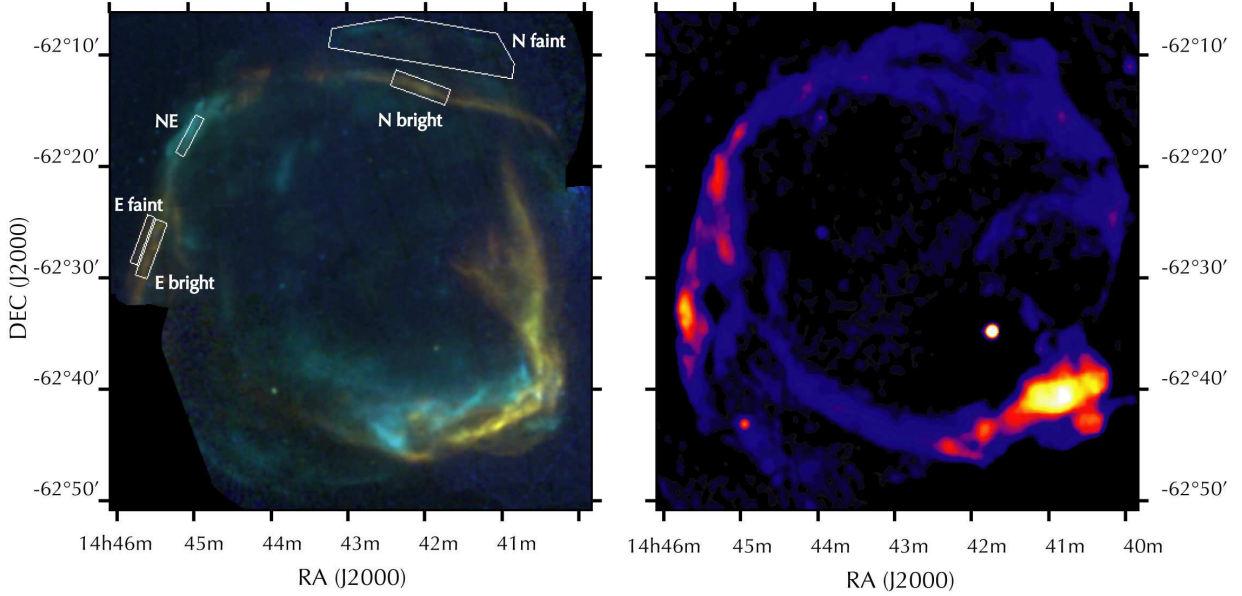,width=0.9\textwidth}
}
\caption{
Left: 
\xmm-EPIC (MOS/PN) mosaic of \rcw, with a color coding similar
to Fig.~\ref{fig_plate} (Plate 1),
using a square root brightness scaling.
Spectral extraction regions are overlayed.
Right: archival MOST 0.84 GHz radio map \citep{whiteoak96,dickel01}.
\label{fig_maps}
}

\end{figure*}

\section{Introduction}

Since the discovery of X-ray synchrotron radiation from SN1006 \citep{koyama95}
it has been found that most young, shell-type supernova remnants (SNRs) 
emit X-ray synchrotron radiation \citep[][for a review]{vink06b}.
For the youngest SNRs Cas~A, Kepler (SN1604) and Tycho (SN1572) this
radiation is confined to
a narrow region close to the shock front. This does not seem to be the
case for somewhat older, but physically much larger 
objects like RCW 86, G266.2-1.2 \citep{slane01a}
and G347.3-0.5 \citep{cassam04c}. The X-ray emission from the 
latter two SNRs does in fact only reveal synchrotron radiation,
and both have been detected in TeV gamma-rays
\citep[][]{aharonian05,aharonian04}.

It has been shown that the thickness of the X-ray synchrotron emitting region
is directly related to the post-shock magnetic field strength 
\citep[e.g][]{vink03a}.
The relatively strong fields found are probably a result
of magnetic field amplification by cosmic ray streaming 
\citep[e.g.][]{bell04,bykov05}.
The presence of X-ray synchrotron radiation is in itself an indication of
efficient acceleration, and if the cut-off energy of the photon spectrum 
is determined by synchrotron losses, it is independent of 
the magnetic field strength, but scales with shock velocity,
as $\propto V_s^2$ \citep{aharonian99}.

Here we report on the analysis of \chandra\ and \xmm\ data of the 
northeastern (NE) part of \rcw\ (G315.4-2.1). 
\rcw\ is an interesting SNR. The non-thermal X-ray emitting regions are broader
than those of the historical SNRs, and not confined to the forward
shock region, possibly as a result of projection effects \citep{vink97}.
In that respect \rcw\ resembles  G266.2-1.2 and G347.3-0.5.
However, unlike those SNRs \rcw\ also emits noticeable
thermal X-ray emission. This allows the determination of
plasma properties, which
can help to determine the conditions that may potentially 
lead to X-ray synchrotron emission in older SNRs.
In particular the NE part 
seems best suited for that 
purpose, since the X-ray synchrotron radiation is confined to the region
directly behind the shock front, and, as a result,
the geometry of the emitting region is easier to asses.

\begin{table*}
\begin{center}
\caption{Best fit models for the \xmm\ MOS1\&2 spectra.}
{\footnotesize
\begin{tabular}{lcccccc}

\hline\hline\noalign{\smallskip}
          &       NE           & E bright      & E faint         & N bright     & N faint\\
\noalign{\smallskip}\hline
\noalign{\smallskip}
$EM{_1}$ ($10^{53}$cm$^{-3}$kpc$^{-2}$) 
               & $12.6\pm1.9$ & $311\pm45$    & $70.6\pm 8.0$      & $219\pm48$        &$69.5$\\
$kT_{\rm e 1}$ (keV) 
               & $6.7\pm2.6$  & $0.57\pm0.05$ & $0.93\pm0.07$   & $0.96\pm0.13$&$6.3\pm1.6$\\
$n_{\rm e}t_1$ ($10^9$~\netunit)
               & $2.25\pm0.15$& $6.7\pm0.6$   & $5.67\pm0.03$   & $4.0\pm0.3$  &$2.27\pm0.12$\\

\noalign{\smallskip}
$EM{_2}$  ($10^{53}$cm$^{-3}$kpc[$^{-2}$)
               & -            &  $43.8\pm5.3$   & -               & $40.9\pm5.6$        &  -\\
$kT_{\rm e 2}$(keV) & -       &  $3.0\pm0.3$  & -               & $3.2\pm0.6$   & -\\
$n_{\rm e}t_2$ ($10^9~$\netunit)& -  & $17.0\pm0.5$  & -               & $19.7\pm0.9$  & -\\
\noalign{\smallskip}
O              & 1           &\multicolumn{2}{c}{$0.56\pm0.06$} & $0.68\pm0.09$ & 1\\
Ne             & 1           &\multicolumn{2}{c}{$0.59\pm0.04$} & $0.72\pm0.07$ & 1\\
Mg             & 1           &\multicolumn{2}{c}{$0.44\pm0.04$} & $0.39\pm0.05$ & 1\\
Si             & 1           &\multicolumn{2}{c}{$0.27\pm0.04$} & $0.38\pm0.07$ & 1\\
Fe             & 1           &\multicolumn{2}{c}{$0.64\pm0.07$} & $0.79\pm0.14$ & 1\\
\noalign{\smallskip}
PL Norm ($10^{-3}$s$^{-1}$keV$^{-1}$ cm$^{-2}$)
               & $1.22\pm0.03$&     -         &   -             &  -            &$1.8\pm0.1$\\
$\Gamma$       & $2.82\pm0.04$&     -         &   -             &  -            &$3.5\pm0.2$\\
$N_{\rm H}$ ($10^{21}$~cm$^{-2}$)  
               & $4.1\pm0.1$ &\multicolumn{2}{c}{$3.6\pm0.2$}   & $3.6\pm0.2$   &$3.9\pm0.2$\\
\noalign{\smallskip}
C-stat/d.o.f.  & 190.6/169  &\multicolumn{2}{c}{(302+260)/230}& 182/85         & 242/103  \\
\noalign{\smallskip}\hline
\end{tabular}
\tablecomments{The models consist of either one or two NEI components, 
or of an NEI component with an additional power law component.
Abudances are given with respect to the solar abunances of \citet{anders89}.
An entry ``1'' means the abundance as fixed to the solar value.
Entries covering two columns were obtained by fitting 
two spectra simultaneously, and forcing
the parameters to be identical. The emission measure, $EM$, is defined as
$\int n_{\rm e}n_{\rm H} dV/d^2$. 
The power law normalization is given at 1 keV.
All errors are 1$\sigma$ errors ($\Delta\chi^2=1$).\label{tab_spectra}
}}
\end{center}
\end{table*}

\begin{figure*}[t]
\centerline{
  \psfig{figure=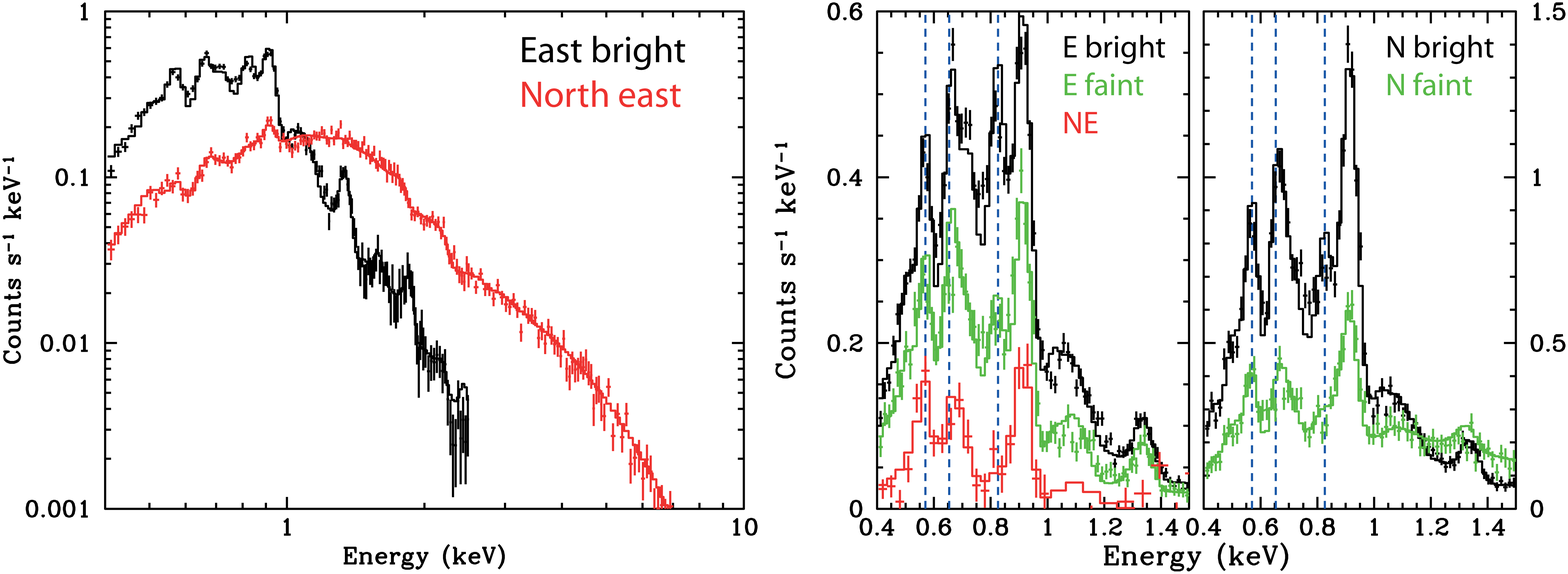,width=0.95\textwidth}
}
\caption{\xmm\ EPIC-MOS spectra from the regions labeled in
Fig.~\ref{fig_maps}.
Left: logarithmic plots of a thermal and a non-thermal spectrum.
Right: a comparison of the line emission from various regions.
From the northeastern spectrum (in red, left panel)
the best fit power law model 
has been subtracted in order to emphasize the thermal emission.
Dashed lines indicate (from left to right)
the energies of OVII He$\alpha$, OVIIILy$\alpha$,
and Fe XVIII line emission.
\label{fig_spectra}
}
\end{figure*}

\section{Observations and Analysis}
\begin{figure*}
  \psfig{figure=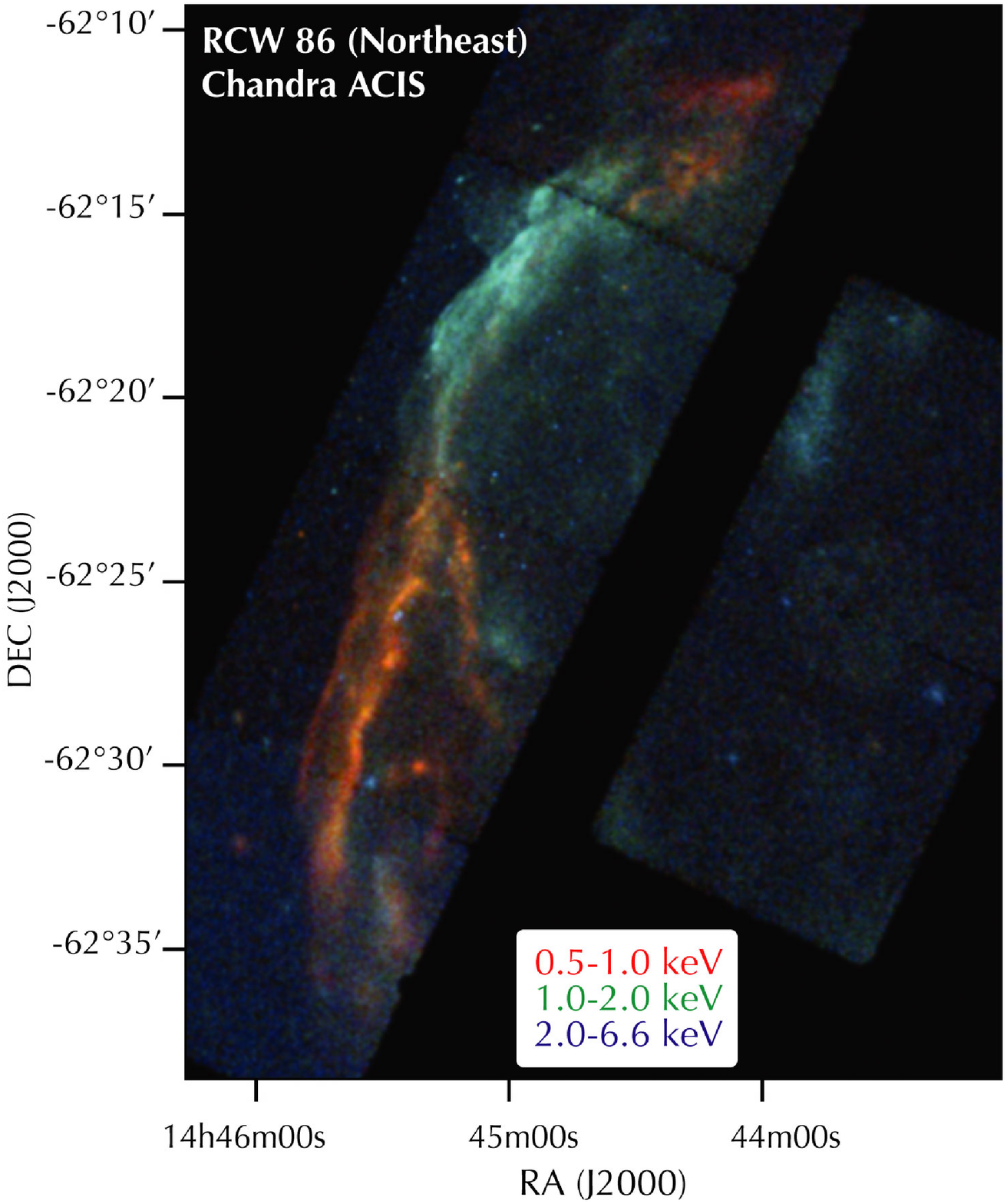,width=0.95\textwidth}
\caption{(Plate 1)\chandra\ exposure corrected 
map of the northeastern part of \rcw, using a square root brightness scaling.
The red,green and blue channels correspond to the
0.5-1 keV 1-1.95 keV and 1.95-6.6 keV energy bands.
\label{fig_plate}
}

\end{figure*}

The NE  of \rcw\ was observed by \chandra\ and \xmm\ for their
joint program.
The \chandra\ observation (ID 500436) has a net exposure time of 72.6 ks  and
was made on June 6, 2004. The \xmm\ observation (ID 0208000101) 
was made on January 26, 2004 with an exposure time of 60 ks. Both \chandra\ 
and \xmm\ observed other parts of \rcw\ before \citep[e.g.][]{rho02}.
Here we present \chandra\ imaging of the NE  part of \rcw\
(Fig.~\ref{fig_plate}/Plate 1), 
but for spectroscopy we focus on the spectra obtained with\
the EPIC-MOS \citep{turner01} instrument of \xmm, 
since it offers the best spectral resolution. For comparison we also
present an analysis of EPIC-MOS data of an observation of the northern part 
of \rcw\ (Obs ID 0110011401, 18 ks). Moreover, we used all six observations
of \rcw\ by \xmm\ to produce the X-ray map shown in 
Fig.~\ref{fig_maps}.
It also shows the spectral extraction regions.
Background spectra were taken from regions outside the SNR.
The data reduction was done with the standard software packages ciao 3.0 for
\chandra\ and SAS 6.5.0 for \xmm.

\begin{figure}[t]
\centerline{
  \psfig{figure=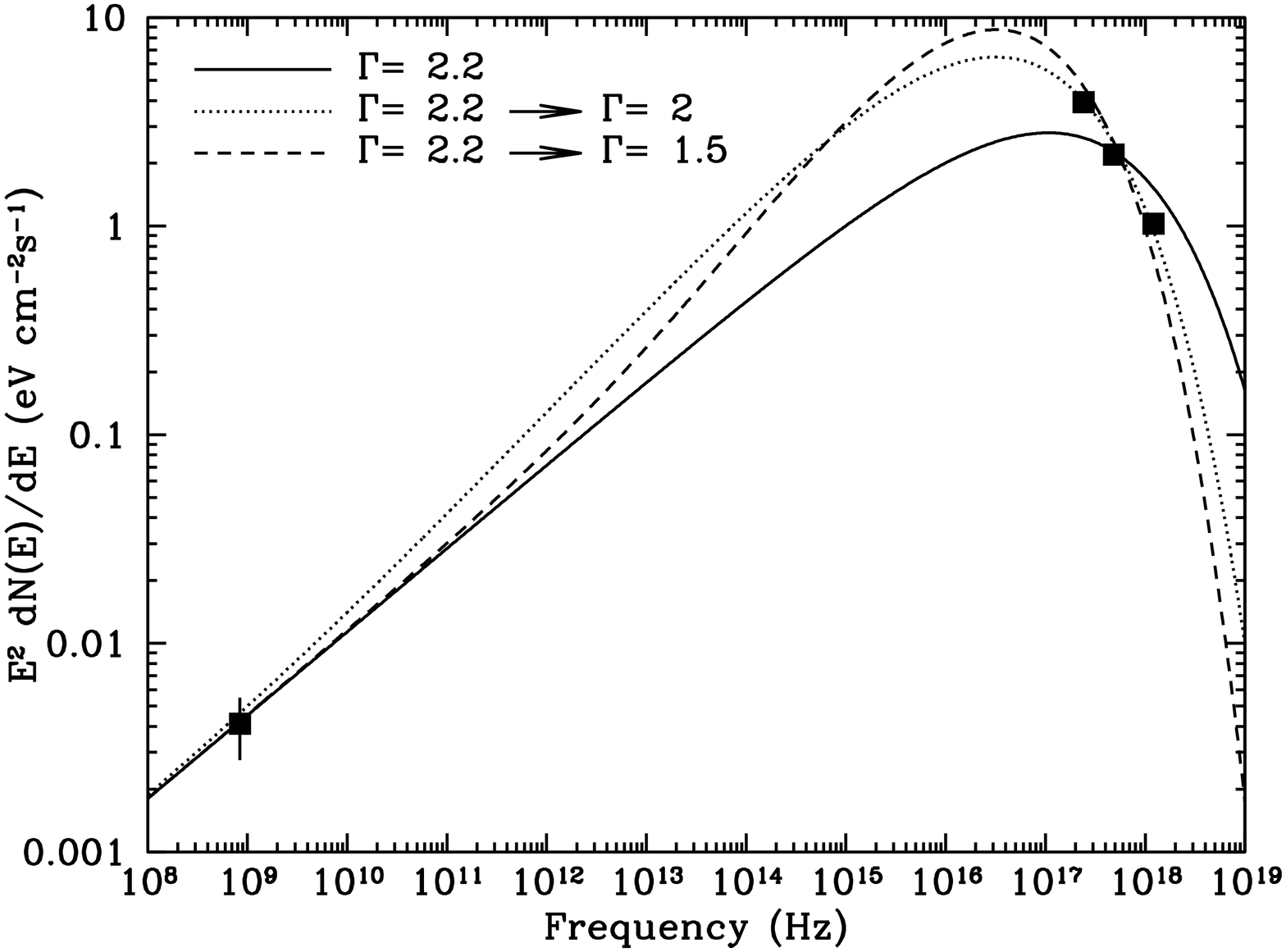,angle=-90,width=0.9\columnwidth}
}
\caption{Broad band ($\nu F_\nu$) 
spectrum of the NE region
with the 847~MHz
flux density based on archival MOST data. The models consist
of synchrotron radiation from power law, and concavely shaped electron energy
distributions with exponential cut-offs at high energies, convolved with the
synchrotron emission function \citep{ginzburg69}. The radio
spectral index is fixed to $\alpha = -0.6$ \citep{caswell75}.
\label{fig_broadband}
}
\end{figure}

The spectrum from the region labeled ``NE'' (Fig.~\ref{fig_spectra})
shows an absence of bright line features.
This is usually taken as a sign for synchrotron emission
\citep{bamba00,borkowski01,rho02}, but for \rcw\
it has been argued by some of us in the past that non-thermal bremsstrahlung
may be viable mechanism as well \citep{vink97}. 
One of the reasons was the detection
of Fe-K emission at 6.4~keV in the southwest coinciding with
the hard X-ray continuum. This interpretation 
was disputed by \citet{rho02}, based on energetic grounds.
Here we concur with that view. In fact the \xmm\ and 
\chandra\ spectra of the NE region do not show any
evidence for Fe-K emission, providing additional evidence for an X-ray
synchrotron interpretation.

Figure~\ref{fig_plate} shows that along the shell
the X-ray emission changes from thermal radiation to predominantly
synchrotron radiation. Unlike in other parts of \rcw, it is 
unambiguous
that in the NE the non-thermal emission starts at the shock front.
Surprisingly, the radio map shows that in the NE 
the X-ray synchrotron radiation is not associated
with bright radio emission, but instead relatively weak radio emission.
In order to better determine the plasma
conditions of the X-ray synchrotron emitting regions,
we do not present here the spectrum extracted from the whole X-ray synchrotron
emitting region, but from a smaller region,
labeled ``NE'' in Fig.~\ref{fig_maps}, which shows more traces  
of thermal emission.
Apart from the NE shell, we also
analyzed a faint region outside the main shell (``N-faint''),
which shows up better in
radio than in X-ray.

We fitted the EPIC-MOS spectra using the spectral analysis software SPEX
\citep{kaastra00}. The spectra are reasonably well described by a combination
of two non-equilibrium ionization (NEI) models or, for the X-ray synchrotron
dominated regions, by one NEI model and a power law spectrum 
(Table~\ref{tab_spectra}). 
The spectral fits indicate extreme departures from collisional equilibration,
even compared to other young SNRs, except for SN1006 
\citep{vink03b}. Although the fit quality is not always satisfactory, both
the fitted models and the raw spectra itself (Fig.~\ref{fig_spectra}) show 
that the fainter the
thermal emission, the lower the ionization degree. 
This can be judged from the increase
of O\,VIII with respect to the O\,VII emission 
and presence of Fe L emission from the brighter parts.
There is also a hint that the X-ray synchrotron emitting regions
contain hotter plasma, and are characterized by lower ionization parameters
(\net).
However, we caution that, since most of the continuum
is non-thermal, the temperature is solely determined by line ratios, making
it more sensitive to potential errors in the atomic data.
This may also explain the peculiarity that whenever a second thermal
component was fitted, its best \net\ value was much higher than the
principal component. The second component 
improved, in particular, the fit between
1 and 1.2 keV, dominated by $n=3\rightarrow 1$\ Ne IX line emission.

\section{Discussion}
A recent development in the interpretation of X-ray synchrotron
radiation is that the width of the region can be used to estimate the 
downstream magnetic field. However, different groups have used, at face value,
different methods for estimating the magnetic field:
\citet{vink03a} assumed that the width, $l$, 
of the X-ray synchrotron emitting region
is determined by a combination of plasma velocity relatively to the shock 
front, $u$, and the synchrotron loss time $\tau_{loss} = 637/B^2E$~, i.e.
$l_{adv} = u \tau_{loss}$, 
with $E$ the particle energy in erg and $B$\ the magnetic
field strength in Gauss.
The other method assumes that the width corresponds to
the diffusion length scale \citep{bamba05,voelk05},
given by $l_{diff} = D/u$, with $D = cE/3eB$ the diffusion coefficient.
In essence  $l_{diff}$\ is the length scale at which advection starts to 
dominate over diffusion as means of transporting particles.
Both methods give a combination of $E$ and $B$, which can be solved by
using the fact that the observed photon energies peak around 
$\epsilon = 7.4 B E^2~{\rm keV}$.
Both methods rely on different assumptions. First of all,
for standard shocks $u = V_s/4$, but for very efficient shock acceleration
the compression factor may be stronger than a factor 4.
Secondly, the diffusion length method assumes  $D = cE/3eB$, which is
the smallest diffusion coefficient possible, as the particle mean free
path is then equal to the gyro-radius (the ``Bohm limit'').

It turns out that with both methods very similar magnetic field estimates
are obtained \citep{ballet05,vink06b}. 
As shown by \citet{vink06b}, this is to be expected
if one observes the X-ray synchrotron spectrum near
the spectral cut-off energies:
The acceleration time for particles according to the first order Fermi 
acceleration theory is, within a factor of order one, 
$\tau_{acc} \approx D/u^2$\ \citep{malkov01}. 
For electrons the acceleration is limited by synchrotron losses.
So there is only a net acceleration if $\tau_{acc} < \tau_{loss}$, 
and the maximum energy is reached when 
$\tau_{acc} \approx \tau_{loss}$ \citep{reynolds98}.
So: 
\begin{eqnarray}
\tau_{acc} \approx \tau_{loss} 
 \Longleftrightarrow 
 D/u \approx u \tau_{loss} & \Longleftrightarrow 
l_{diff} \approx l_{adv} 
\end{eqnarray}
Therefore, $l_{diff} \approx l_{adv}$\ is the geometrical equivalent of
$\tau_{loss} \approx \tau_{acc}$.
Note that the {\em observational} fact that the diffusion length scale and 
advection length scale methods give similar results is a justification
for the assumption that the diffusion coefficient is close to the  Bohm limit,
and a compression ratio close to the standard value of 4 \citep{vink06b}.

The \chandra\ image reveals a width of the X-ray synchrotron shell of
$\sim100$\arcsec, corresponding to $3.7\times10^{18}$~cm
for a distance of $d=2.5$~kpc 
\citep{westerlund69,rosado94}. 
Fitting a projected shell model we
estimate a physical width of $1.7\times10^{18}$~cm.
In addition we apply a factor 0.6, because the actual width
is a convolution of advection and diffusion processes (i.e. we
set $l_{diff} = l_{adv}$ in Eq. (1) of \cite*{berezhko04a}),
so 
$l_{adv} = 1.0\times10^{18}$~cm.
Assuming that the 
shock velocity is $V_s=600$~\kms\ \citep{ghavamian01} 
gives inconsistent results for the two methods:
$B\sim90$~\mug\ 
employing the diffusion length method, and 6~\mug\ assuming
the advection length method. 

However, if we no longer assume a shock velocity of 600~\kms,
which after all was based on optical observations of a different part
of \rcw\ from which no X-ray synchrotron radiation is emitted, we
can use the condition that $l_{adv} = l_{diff}$ in order to
estimate $V_s$ and $B$,
using an observed photon energy of $\epsilon \approx 1$~keV.
We found that $B$
can be directly estimated from the 
diffusion/advection length alone, whereas the plasma velocity only depends
on the photon energy:
\begin{eqnarray}
B \approx \bigl(\frac{c}{3e}\bigr)^{1/3}l_{adv}^{-2/3} = 
24 \bigl(\frac{l_{adv}}{1.0\cdot 10^{18} {\rm cm}}\bigr)^{-2/3}\ 
{\rm \mu G},\\
V_s = u \chi \approx \chi \sqrt{\frac{\epsilon}{7.4}
\frac{c}{3e \cdot 637} }
= 2650 \cdot \frac{\chi}{4} \sqrt{\frac{\epsilon}{1\, {\rm keV}}}\,  
{\rm km/s}, \label{eq_aharonian}
\end{eqnarray}
where $\chi$ the shock compression factor. 
Eq.~(\ref{eq_aharonian}) was previously reported by \citet{aharonian99}.
To get more accurate estimations one has to solve the, model dependent, 
kinetic equations for the particle distribution.
For the moment, we estimate the error in the physical width due to distance
uncertainties and projection effects to be $\sim 30$\%,
resulting in an error in $B$ of $\sim 5$~\mug.
The downstream magnetic field
therefore appears to be lower than for other SNRs \citep{vink06b}.
There is some uncertainty in the actual cut-off energy of the synchrotron
spectrum (see below), but uncertainties about the assumptions -
larger compression ratio, less efficient diffusion -
make Eq.~(\ref{eq_aharonian}) effectively a lower limit.

One may wonder why the shock velocity in some
regions may be so much  higher, and why it has not resulted in
a more distorted shell.
Here the thermal spectra helps to answer this question.
For all fitted regions the  \net\ value of the primary component
is low, being \net$=6.7\times 10^9$~\netunit\ even for the brightest region.
From the size of our extraction box,
we estimate that the emitting volume is $V= 10^{56}-10^{57}$~cm$^3$, 
together with the emission measure this implies an electron density of 
$n_{\rm e}= 0.5-1.6$~cm$^{-3}$. If we use this to estimate how long
ago the plasma was shock heated we find $t\lesssim 425$~yr. 
This is surprisingly short for a large SNR as \rcw.
In such a time
the difference in radius between regions with low and high shock velocity
would be $\sim 0.9$~pc on an average shock radius of 16~pc.
The short interaction time supports the idea that
\rcw\ is a SNR expanding in a wind
blown bubble \citep{vink97}. Such SNRs expand rapidly for a long
time, but the shock velocity drops rather suddenly as soon as the shock
starts interacting with the surrounding shell swept up by the stellar wind
\citep[e.g.][]{dwarkadas05}. 
This suggests that in \rcw\ the shock
has reached in some regions the dense shell around the bubble some 400~yr ago,
after which it rapidly decelerated. 
In other regions the shock velocity is
still high, but, due to the low density, the thermal X-ray emission is weak.
This explains the coexistence of relatively weak radio synchrotron
emission with conspicuous X-ray synchrotron emission:
due to the lower density fewer electrons
are accelerated, but because of the high shock velocity they
can be accelerated to higher energies.

Interestingly, the X-ray synchrotron radiation is relatively bright,
with respect to the radio emission,
because we find that the simplest broad band synchrotron model, i.e.
synchrotron radiation from a power law electron spectrum with an exponential
cut-off does not fit the data (Fig.~\ref{fig_broadband}).
It can explain the X-ray flux, but not the
spectral slope. Instead the electron spectrum needs to be concave,
as predicted by non-linear shock acceleration models.
However,
we can not determine whether the electron power law index bends toward -2,
predicted for an overall shock compression ratio of 4, or toward -1.5,
predicted for strongly cosmic ray modified shocks
\citep{berezhko99}.

The NE region of \rcw\ has 
properties resembling
those of the TeV emitting SNRs G347.3-0.5,
and G266.2-1.2: weak radio emission, and X-ray emission (almost)
entirely consisting of synchrotron radiation. 
For \rcw\ the (weak) thermal X-ray emission indicates that these properties
are due to a low density combined with a, relatively, high shock velocity.
We speculate that  for G347.3-0.5 and G266.2-1.2 the shock
also moves through a low density region, e.g. a stellar wind bubble,
and 
the shock velocity is similarly high. 
\rcw\ may be different in that
some parts of the shock have reached the shell.
Inside the bubble the shock evolution can be approximated by the Sedov
self-similar model, but this breaks down as soon as the shell is reached.
Finally, 
assuming a Sedov 
evolution and using the apparent radius of 22\arcmin,
the age of the remnant is estimated  to be
$t = \frac{2r}{5V_s} \approx 2250 (V_s/2700\, {\rm km\,s^{-1}})$~yr.
This would put the explosion date of \rcw\ closer to AD 185, 
the year a putative supernova was observed in China \citep{stephenson02}.
A shock velocity of $\sim 600$~\kms\ 
would be more consistent with a 10,000~yr old SNR. 
Our results, therefore, strengthen 
the case that the event recorded
by Chinese astronomers was indeed a supernova and that \rcw\ is its remnant.

\acknowledgements
JV is supported by a Vidi grant
from the Netherlands Organisation
for Scientific Research (NWO).

\end{document}